# Quantum Linear Magnetoresistance and Fermi Liquid Behavior in Kagome Metal Ni$_3$In$_2$S$_2$

*P. Das, P. Saha, M. Singh, P. Kumar, and S. Patnaik*

[1] School of Physical Sciences, Jawaharlal Nehru University, New Delhi-110067, India

Corresponding author: spatnaik@jnu.ac.in

Kagome metals gain attention as they manifest a spectrum of quantum phenomena, including superconductivity, charge order, frustrated magnetism, and intertwined correlated states of condensed matter. With regard to electronic band structure, several of the them exhibit non-trivial topological characteristics. Here, we present a thorough investigation on the growth and the physical properties of single crystals of Ni$_3$In$_2$S$_2$ which is established to be a Dirac nodal line Kagome metal. Extensive characterization is attained through temperature and field-dependent resistivity, angle-dependent magnetoresistance and specific heat measurements. In most metals, the Fermi liquid behaviour is mostly restricted to a narrow range of temperature. In Ni$_3$In$_2$S$_2$, this characteristic feature has been observed for an extensive temperature range of 82 K. This is attributed to the strong electron-electron correlation in the material. Specific heat measurements reveal a high Kadowaki-Woods ratio which is in good agreement with strongly correlated systems. Almost linear positive magnetoresistance follows the conventional Kohler scaling which depicts the applicability of semi-classical theories. The angle-dependent magneto-resistance been explained using the Voigt-Thomson formula. Furthermore, de-Haas van Alphen oscillations are observed in magnetization vs. magnetic field measurement which sheds light on the topological features in the Shandite Ni$_3$In$_2$S$_2$.

## 1. Introduction

Topological semimetals (TSMs) with distinct quasi-particle transport characteristics, such as Dirac, Weyl, nodal-line, and nodal-chain dispersion, have garnered significant current interest in condensed-matter systems. The underlying physics associated with such material realization of field-theory principles is further accentuated in low-dimensions due to nontrivial electronic band topology [1-4]. Of substantial current interest is the group of materials under the gamut of "Kagome metal" that have proliferated as a new topological material class [5-9]. Several recent publications have identified geometrically frustrated Kagome systems as promising candidates for next-generation optoelectronics, and spintronics applications [10-12].

In general, Kagome compounds with their corner-shared triangle networks for electron transport, generate intricate quantum interference processes giving rise to an admixture of flat bands, saddle points, and van Hove singularities (VHSs) [13-15]. Of particular interest is $Ni_3In_2S_2$ which belongs to the family of Shandite structured Kagome-metal, which requires in-depth study for clarification because the material that is currently available has inconsistent and contradicting information. From density-functional theory (DFT) calculations and symmetry-based theories, $Ni_3In_2S_2$ is determined to be a nonmagnetic topological semimetal with six endless Dirac-nodal lines near the Fermi level [16-17]. Earlier reports have claimed that this non-magnetic Kagome metal exhibits a high magnetoresistance of 2000% which is attributed to the high mobility and small effective mass of the conduction electrons along with de Haas van Alphen (dHvA) oscillation and Wilson ratio ($R_w$) greater than unity, suggesting may have strong correlation effects in Ni3In2S2 [16]. Another study suggests that $Ni_3In_2S_2$ is a completely compensated semimetal with record-high mobility, and extremely high magnetoresistance [17]. However, the physical mechanism underpinning the electrical conduction in $Ni_3In_2S_2$ needs more clarity vis a vis its band structure. In this work, we provide additional evidence for strong electron correlation by means of corroborative heat capacity measurements, supporting the notion that correlation effects are prominent in $Ni_3In_2S_2$. The role played by electron correlation, which is the basis for charge density wave [18-20] and superconductivity [21-23] in analogous systems, also needs to be integrated for the explanation of transport data. It is well known that the electron-electron interaction strongly influences the transport and thermodynamic characteristics of a metallic system. For instance, a low temperature $T^2$ behavior can be observed in the temperature-dependent resistivity in good metals which is ascribed to electron-electron interaction. The typical behaviour is often called the Baber law [24-26]. Generally, in metallic systems, this $T^2$ nature exists only till low temperatures (below 20K) [27]. At high temperatures, electron-phonon scattering dominates over the electron-electron scattering, which ultimately leads to a linear T or $T^5$ dependence. However, recent studies on dichalcogenides like $TiTe_2$ [28] and $TiS_2$ [29] have revealed the presence of electron-electron interaction at temperatures as high as 400K. But, the reason behind this unconventional behaviour still requires in-depth analysis. Furthermore, it is obvious to inquire whether such unusual behaviour exists in other materials such as the Kagome dichalogenides.

In this paper, we report on the synthesis and extensive magneto-transport and thermodynamic characterization of single crystals of $Ni_3In_2S_2$. We observe a quadratic temperature dependence of resistivity ($T^2$) in a broad temperature window. In the high-temperature region, on the other hand, a linear dependence on temperature is observed. Furthermore, the specific heat study reconfirms the existence of strong electron-electron interaction in a wide range of temperatures. The Kadowaki-Woods ratio, which is a measure of strength of electronic correlation, is found to be comparable to strongly correlated systems such as $MoOCl_2$, $Na_{0.7}CoO_2$ and many other heavy fermionic systems [30-32]. The linear positive magnetoresistance of $Ni_3In_2S_2$ is 286% at 2K, following the standard Kohler equation, shedding light on the semi-classical origin elevated magnetoresistance. This type of non-saturating MR could be associated with high carrier concentration [33], carrier spin polarization [34], or extremely high mobility [35-36]. We further validate the semi-classical nature of Magnetoresistance in our studied material by elucidating the angle-dependent Magnetoresistance using the Voigt-Thomson formula. Furthermore, in the magnetic susceptibility data, we found faint but noticeable oscillations i.e., de Haas van Alphen oscillations (dHvA). This result adds more evidence to support the sample's intrinsic topological nature.

## 2. Experimental Techniques

### 2.1. Sample Synthesis

Single crystals of $Ni_3In_2S_2$ were synthesized using a modified Bridgemann furnace. Nickel (Powder, Aldrich, 99.99%), Indium (Powder, Thermo scientific, 99.999% pure) and Sulfur

(Powder, Aldrich, 99.98% pure) were taken in stoichiometric ratio and then ground using mortar and pestle. The powder mixture was sealed in a quartz tube under the pressure of $10^{-2}$ mbar and then fired at 1000°C for 24hrs followed by cooling down to 800°C over 100 hrs. Finally, the ampule was water quenched at 800°C. Shiny silver crystal flakes were obtained. About four crystals were chosen for transport measurements, the reported data in this manuscript were taken on a $2.1 \times 1.8 \times 0.4$ mm³ sized single crystal. The electrical contacts between the sample and copper wires were made using conductive silver epoxy, which was dried using heat treatment. For the specific heat measurements, a 35mg crystal of size $1.24 \times 0.96 \times 0.38$ mm³ was used. Apiezon N grease was used to attach to the sample to the platform.

*2.2. Structural characterization*

The crystal phase and structure of the sample were identified using powder x-ray diffraction (XRD) in Rigaku Miniflex D/MAX 25000PC (Cu-Kα radiation, λ=1.5460Å) at room temperature. Figure 1(a) shows the XRD data of single crystal of $Ni_3In_2S_2$ which was cleaved along (0 0 l) plane. This Shandite Kagome-metal belongs to space group $R\bar{3}m$ (group no. 166). The lattice parameters from the refined XRD data are: a=b=5.37Å and c=13.53Å, which is in agreement with previous reports [16-17]. Evidently, the space group and lattice parameters of $Ni_3In_2S_2$ and $Co_3Sn_2S_2$ are much close to one another [37]. The conventional unit cell of $Ni_3In_2S_2$ is shown in the figure (1b). Inset (b) shows the single-XRD pattern i.e., Laue's pattern, which reconfirm the phase purity of the crystal with the powder-XRD data.

*2.3. Physical Property Measurements*

The temperature-dependent resistivity and high-field magnetoresistance were performed in *Cryogenic* Cryogen Free Magnet (CFM, 1.6K, 8T) in conjunction with variable temperature insert. The standard four-probe method was used for resistivity measurements. The magnetization measurement and specific heat measurement was carried out using Physical Property Measurement system (PPMS, 1.8K, 14T).

## 3. Result and Discussions

### 3.1. Electrical Resistivity

The temperature (T) dependence of electrical resistivity at zero external magnetic fields is shown in Figure (2). The increasing resistivity with evolving temperature indicates the typical metallic behaviour of the sample in the broad temperature range of 3.6 -300K. The residual resistivity ratio RRR is estimated to be 42.4. No abrupt change in slope is observed in RT data which implies the absence of any magnetic phase transition in this compound. The curve is fitted with the resistivity equation $\rho(T) = \rho_{res} + \rho_{e-e}(T) + \rho_{e-ph}(T)$; where the residual resistivity $\rho_{res} = 0.67013\ \mu\Omega.\,cm$, $\rho_{e-e}(T) \sim T^2$ and $\rho_{e-ph}(T) \sim T$. According to the fitting analysis in Fig. (2), resistivity exhibits a $T^2$ dependence at low temperatures, that is sustained to a remarkably high temperature of 86K (light orange region). This points towards the Fermi liquid behaviour of the sample. It is to be noted that this Fermi liquid behaviour also known as Baber law [24], can be confirmed through the specific heat study as will be explained in the next section. The resistivity varies at low temperature region as $\rho(T) = \rho_{res} + AT^2$ [17], where $\rho_{res}$ is the residual resistivity and coefficient $A$ denotes the electron-electron scattering effects and is typically proportional to the square inverse of Fermi energy i.e., $A \propto 1/\varepsilon_F^2$ [38]. The value of $A$ obtained from the fitting of resistivity data is $0.726 n\Omega.\,cm.\,K^{-2}$. This finding suggests that for transport in $Ni_3In_2S_2$, electron-electron scattering dominates over a broad temperature range. The presence of Fermi liquid behaviour in such a broad temperature range is has not been observed in the previously reported Shandite compounds. Above this temperature (82K), the temperature dependence of resistivity switches to a linear T dependency (light blue zone). In essence, it is observed that in $Ni_3In_2S_2$, the resistivity temperature scaling switches from $T^2$ to $T$ with a small crossover region (light green). The temperature dependence in the small cross over region is found to be $\sim T^{3/2}$. Similar small crossover sections have also been observed in layered compounds [30], where the Baber law holds at low temperatures followed by linear behaviour at higher temperatures. The linear temperature dependence of resistivity above 108K is reflective of dominance of electron-phonon scattering similar to what is observed in Shandite family $Co_3In_2S_2$ [39].

## 3.2. Specific Heat

Additionally, heat capacity measurements have been carried out to verify the linear temperature dependence of specific heat at low temperatures which is another determining feature of Fermi liquids. The specific heat of a metal at low temperature is given by $C = \gamma T + \beta T^3$[40] where $\gamma$ is the Sommerfeld coefficient which indicates the electronic contribution, and $\beta$ is the phononic contribution. In order to probe the electronic contributions to specific heat, one can study the Kadawaki-Woods scaling in fermi liquid systems [34]. In Figure (3) we show the measured specific heat data for single crystalline $Ni_3In_2S_2$. From the previously mentioned specific heat equation, the value of $\gamma$ and $\beta$ obtained are as follows: $\gamma = 4.81 mJ.mol^{-1}.K^{-2}$ and $\beta = 1.7 mJ.mol^{-1}.K^{-4}$. In single band free electron model, $\gamma$ can be related with the effective mass of electrons with the relation: $\gamma = m^* n^{\frac{1}{3}} \left(\frac{k_B}{\hbar}\right)^2 \left(\frac{\pi}{3}\right)^{\frac{2}{3}}$, from which the value of effective mass can be obtained as $m^* = 1.17 m_e$. It is worth noting that the effective mass obtained for this sample is much higher than the others compounds in the same family [37,*]. Previously T. T Zhang et al [15] had reported a high Wilson ratio () for the same compound. This high value indicates towards the strong electronic correlation and is in resonance with our results of specific heat measurement. The phononic contribution parameter β, is connected with the Debye temperature $\Theta_D$. The Debye temperature for $Ni_3In_2S_2$ is estimated to be about 92.09K, which is quite high as compared to other shandite materials.

Now from Figure (3) we have demonstrated that in $Ni_3In_2S_2$, the resistivity temperature scaling directly transitions from $T^2$ to T. The above-mentioned predicted Debye temperature aids us in comprehending this unusual trait. Below the cross-over temperature, the transport is dominated by electron-electron scattering. We note that in $Ni_3In_2S_2$, crossover temperature and Debye temperature are comparable to each other.

The A coefficient and $\gamma$ coefficient obtained from the scaling relation for resistivity curve and specific heat, both contain electron-electron scattering effects. The Kadawaki-Woods ratio $\alpha = A/\gamma^2$ is a significant metric that investigates the relation of the electron-electron interaction rate and the effective mass [41]. Usually, the value of the ratio is approximately same for materials belonging to the same class. While for many transition metals, $\alpha_{TM} \sim 0.4 \; \mu\Omega.cm.mol^2 K^2 J^{-2}$, whereas for many heavy fermionic metals $\alpha_{HF} \sim 10\ \mu\Omega.cm.mol^2 K^2 J^{-2}$. Here, based on the outcomes of our experiments, we obtained $\alpha_{Ni_3In_2S_2} \sim 31.37\ \mu\Omega.cm.mol^2 K^2 J^{-2}$ which is three times higher than the value for heavy fermionic compounds. Such high Kadawaki-Woods ratio has been previously obtained many oxide perovskites [31, 42-43]. This high value of α attributes to the strong electron correlation and is consistent with the broad $T^2$ scaling in the temperature dependent resistivity behaviour.

### 3.3. Magnetoresistance

The field dependence transverse magnetoresistance MR at constant temperature is shown in figure (3). Magnetoresistance can be calculated as: $\mathrm{MR(B)} = \left[(\rho(B) - \rho(B=0))/\rho(B=0)\right] \times 100\%$, where $\rho(B)$ and $\rho(B=0)$ being the resistivity under the applied fixed magnetic field and zero magnetic field respectively. The direction of applied current and the magnetic field is perpendicular to each other (I ⊥ B).

The sample exhibited a notable peak in positive Magnetoresistance (MR), reaching a maximum value of 286% at 2K under a 6T magnetic field. It is noteworthy that our experimental MR value is comparatively lower than the reported results on Ni3In2S2, wherein the MR reaches an exceptionally high value of 15518% at 2K and 13T magnetic field [17]. This discrepancy in MR values is anticipated, attributable to lower carrier mobilities and the absence of effective charge compensation within our sample. According to our findings, the mobility in the Ni3In2S2 single crystal at 2K is $\sim 30.63\ cm^2.V.s$, and the carrier density is $\sim 1.72 \times 10^{19} cm^{-3}$. The MR% remarkably decreases with increasing temperature as shown in inset of fig. 3(a). As the MR varies with the direction of applied magnetic field (anisotropy in MR), the exixtance of open orbit in Fermi surface can be a reasonable explanation for the non-saturating behaviour of MR [45]. In Dirac semimetals, this kind of linear MR originated due to coupling of electrons with the spin-orbit interaction of the material [46-48].

Futhermore, the application of the Kohler rule properly captures the analysis of transverse magnetoresistance (MR) and illuminates the complex mechanisms governing the scattering dynamics of charge carriers. The Kohler rule is defined by a basic equation (1) using the principles of semi-classical band theory, providing a more sophisticated explanation of the intricate processes underlying the observed MR behavior [49]. The Kohlr scaling relation is formulated as follows:

$$MR = \frac{\Delta\rho}{\rho_0} = \alpha \left(\frac{B}{\rho_0}\right)^m \qquad (2)$$

Here, in equation (2), the parameters (α, m) are predefined constants, and $\rho_0$ represents the zero-field resistivity. Adhering to this Kohler rule, the MR data is systematically represented as MR~ $\frac{B}{\rho_0}$ curve at different temperatures and is shown in figure (b). In accordance with Kohler's rule, the MR curve convergence into a single line denotes the homogeneity in the charge carrier scattering process throughout the system's temperature range. The parameter 'm' derived from fitting equation (1), exhibits a value of 1.05 at 2K, indicative of an almost linear dependence of Magnetoresistance (MR) on the applied magnetic field. Consequently, the value of 'm' increases noticeably as the temperature rises. The quantum linear MR (QLMR) is pronounced in Dirac semimetals due to their unique band structure, and the presence of Dirac points such QLMR of Dirac semimetals has been studied extensively in recent past years and is an important tool for studying the electronic and transport properties of these materials. However, Fang, Hongwei, et al found the parabolic dependence of MR on applied magnetic field which suggest this compound as fully compensated semimetal [17].

Additionally, the angular dependence of MR has been studied. Figure (4) shows the angle-dependent magnetotransport characteristics, where Θ is the angle between direction of the magnetic field and the current, as shown in the inset of Figure 4(a). Figure 4(b) shows the $R_{xx}$ as a function of angle Θ at different temperatures at fixed magnetic field B=4T. It is observed that at Θ=90° and 270°, the resistance value reach at its highest point, and at Θ=0°, 180° and 360° the resisistance get minimum. The anisotropic orbital magnetoresistance of a system is defined by the Voigt-Thomson formula [44-45].

$$R_{xx}(\theta) = R_\perp sin^2\vartheta + R_\parallel cos^2\vartheta$$

where $R_\perp = R_{xx}(\vartheta = 90°)$ and $R_\parallel = R_{xx}(\vartheta = 0°)$. The red solid curve in Figure (4) shows that the Voigt-Thomson formula that provides a good fit with the measured values of resisistance

$R_{xx}(\theta)$. This outcome further confirms the semi-classical MR origin in our sample, that relate to the extent of Lorentz force in this measurement configuration.

### 3.4. Magnetization

Quantum oscillations at low-temperature and high magnetic fields is a very useful tool to probe the electronic properties of materials such as determining the Fermi surface of material, effective mass of charge carriers. Figure (5) shows the isothermal magnetization measured with an applied external magnetic field up to 10T H || c for our sample. The de Haas-van Alphen (dHvA) oscillations were present when magnetic field exceeding 7T at 2K, indicating the fermi pockets associated with dHvA oscillations. Moderate oscillations with amplitude 0.046 emu/g at 2K were seen. The major oscillation frequencies $F_\alpha$=12.02T, $F_\beta$=19.24T and $F_\gamma$=28.85T were derived from the fast Fourier transform (FFT) analysis of this oscillatory magnetization as shown in figure 5. The cross-sectional area ($A_F$) of Fermi surface can be calculated by assuming the circular cross-section of surface along (0 0 l) plane using Onsager relation $A_F = \frac{2\pi e}{\hbar} F$, where F is the frequency of the oscillation. The cross-sectional area of Fermi surfaces is $1.15\times10^{-3}\text{Å}^{-2}$, $1.18\times10^{-3}\text{Å}^{-2}$ and $2.76\times10^{-3}\text{Å}^{-2}$ corresponding to the frequencies $F_\alpha$, $F_\beta$ and $F_\gamma$ respectively. The obtained surface area values are used further to calculate the Fermi vectors are as follows $K_\alpha=1.91\times10^{-2}\text{Å}^{-1}$, $K_\beta=1.94\times10^{-2}\text{Å}^{-1}$ and $K_\gamma=2.96\times10^{-2}\text{Å}^{-1}$ respectively using the formula $K_F = \sqrt{\frac{A_F}{\pi}}$.

This dHvA oscillations can be described using Lifshitz-Kosevich (LK) formula [53] in which more detailed information about the Berry phase can be extracted by the fitting of LK formula as given bellow:

$$\Delta m \propto R_T R_D R_S \sin\left[2\pi\left(\frac{F}{\mu_0 B} - \gamma - \delta\right)\right]$$

Where $R_T = \frac{2\pi^2 k_B m^* T / \hbar e \mu_0 H}{\sin\left(2\pi^2 k_B m^* T / \hbar e \mu_0 H\right)}$ due to temperature effect, $R_D = \exp\left(2\pi^2 k_B m^* T / \hbar e \mu_0 H\right)$ due to impurity scattering, and $R_S = \cos(\pi g m^*/2m_e)$ due to spin-splitting effect. The oscillation of magnetization is due to the sine term as mentioned in the equation. The phase term $-\gamma - \delta$, in which $\gamma = \frac{1}{2} - \frac{\Phi_B}{2\pi}$ is called Onsager phase directly linked with Berry phase $\Phi_B$. The other term $\delta$ is additional phase shift, can be determined by the dimensionality of the fermi surface. $\delta = 0$ or $\pm \frac{1}{8}$ for 2D and 3D cases where + is for the electronlike cases and – for the holelike cases.

## 4. Conclusion

In summary, we have successfully grown high-quality single crystals of $Ni_3In_2S_2$, which is a Dirac nodal line Kagome metal. Fermi-liquid behavior up to 86 K, which is the highest among the existing Shandite Kagome semimetals, is observed. This is corroborated by specific heat studies within the corresponding temperature range. The Kadowaki-Woods ratio of $Ni_3In_2S_2$ is in good agreement with the values of strongly correlated metals. Almost linear, unsaturated magnetoresistance as well as the anisotropy in MR are attributed to the existence of open orbits in the Fermi surface of $Ni_3In_2S_2$. Employing the Kohler scaling to magneto-resistance data we checked the applicability of semi-classical theories to magneto-transport in $Ni_3In_2S_2$. This is further complemented by the implementation of the Voigt-Thomson formula that is well-fitted in the angle-dependent magnetoresistance. The observation of de Haas van Alphen oscillations at 2K substantiates the quantum behavior of $Ni_3In_2S_2$. Our findings offer insight into the interplay between electron correlation and topological band structure in Kagome metals, setting the stage for deeper exploration and comprehension of these intriguing materials.

## 5. Acknowledgement

P. Das and P. Saha acknowledge UGC-NET JRF for financial support. M. Singh and P. Kumar thank CSIR for providing JRF. We are grateful to the FIST program of the Department of Science and Technology, Government of India for the use of the low-temperature high magnetic field measurement facility at JNU. We acknowledge funding support from DST towards the procurement of chemicals and consumables from the project (DST/NM/TUE/QM-10/2109(G)/6). We acknowledge Advanced Instrumentation Research Facility (AIRF), JNU for use of the PPMS measurement facility.

**Figure Captions**

Figure 1: Structural characterization of $Ni_3In_2S_2$ single crystal: Figure (a) shows the XRD pattern of $Ni_3In_2S_2$ single crystal cleaved along (0 0 1) plane. Inset (i) shows the typical image of single crystal. Inset (ii) shows the Lau's pattern of single crystal. (b) Schematic view of unit cell of $Ni_3In_2S_2$ with $R\bar{3}m$ symmetry. (c) SEM image of the single crystal shows the layered structure of $Ni_3In_2S_2$.

Figure 2: Temperature-dependence of longitudinal electrical resistivity of $Ni_3In_2S_2$ measured in ab-plane without external magnetic field.

Figure 3: The plot of $C_p/T$ vs. $T^2$ with a linear fitting at a low temperature of 12K by $C_p/T = \beta T^2 + \gamma$ in solid line (red). The inset of figure 3 shows the temperature dependence of specific heat of the single crystal $Ni_3In_2S_2$ in the temperature range 2-160K.

Figure 4: (a) The magnetic field dependence of transverse resistance at different temperature. Transverse magnetoresistance $MR(\%) = [R(H) - R(0)]/R(0) \times 100$ measured at the indicated temperatures. Inset of 4(a) shows the variation of MR with respect of temperature. (b) The MR~$B/\rho_0$ plot data was systematically fitted using Kohler's rule in the relevant equation (1) at different temperatures.

Figure 5: Magneto-transport behavior of Ni3In2S2 single crystal. (a) The angular dependence of $R_{xx}$ at fixed temperature 5K and magnetic field 4T. The solid curve (red) is the fitting using Voigt-Thomson Formula. The inset of (a) clarifies the angle Θ. (b) The angular dependency of $\rho_{xx}$ at different temperatures at fixed magnetic field of 4T. Inset (i) shows the variation of $R_\perp = R_{xx}(\vartheta = 90°)$ and $R_\parallel = R_{xx}(\vartheta = 0°)$ with temperatures. Inset (ii) shows the variation of anisotropy with temperatures.

Figure (6): The dHvA oscillations in $Ni_3In_2S_2$. (a) Isothermal magnetization data under $H \parallel c$ at 2K is shown in inset of (a). Zoom in view of magnetization data from 8-8.4T. (b) The magnetization oscillation after background substruction is shown in inset of (b). The corresponding Fourier transforming of the oscillating component.

**Figure 1**

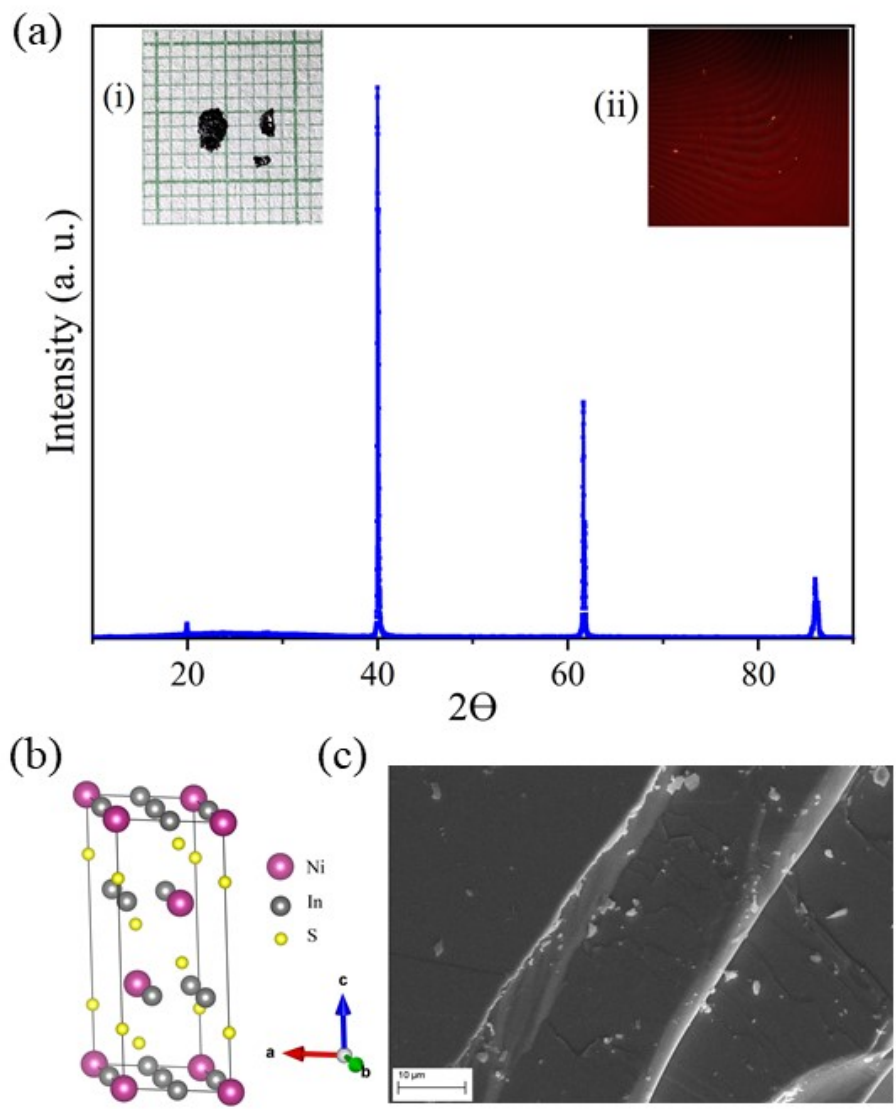

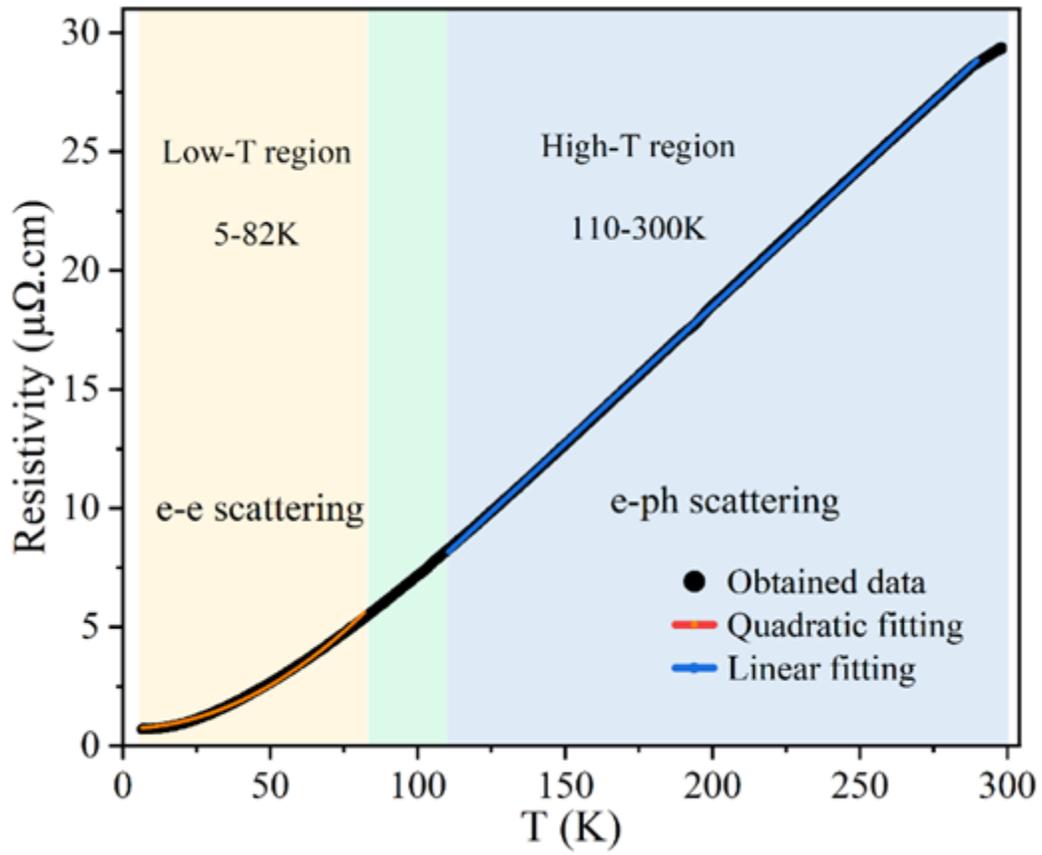

**Figure 2**

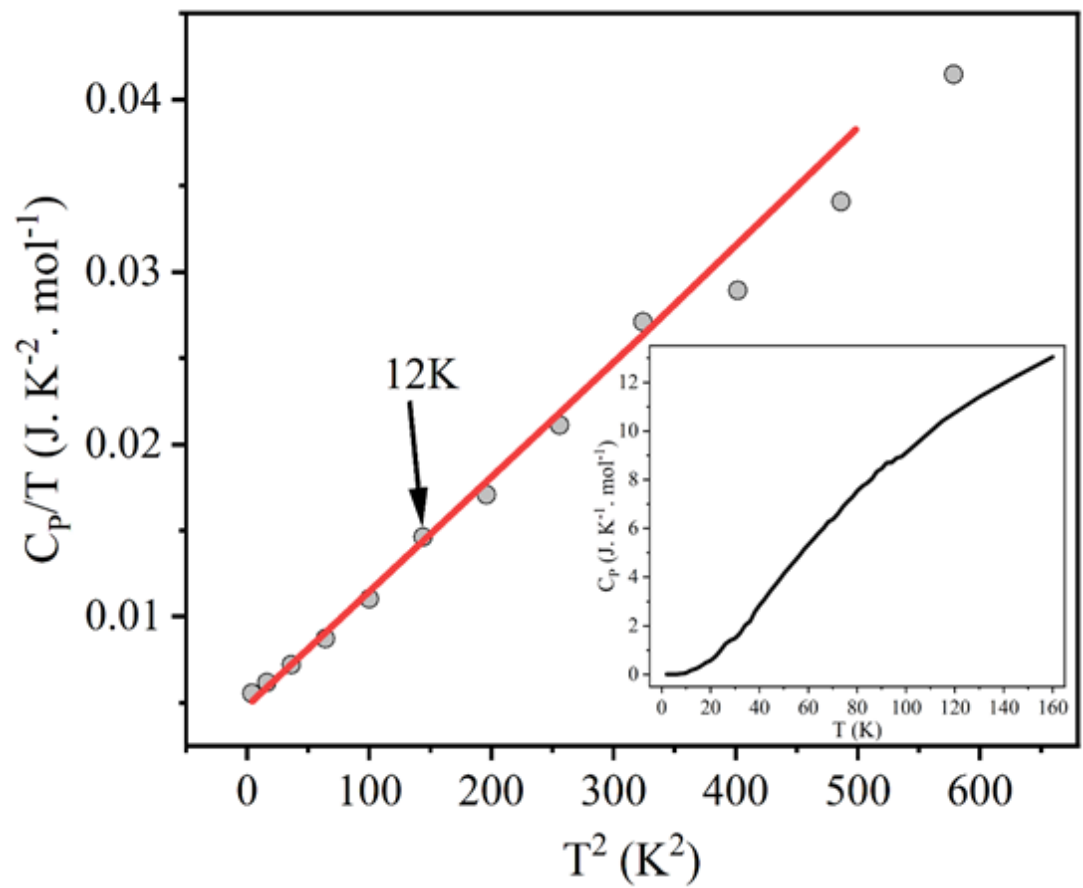

Figure 3

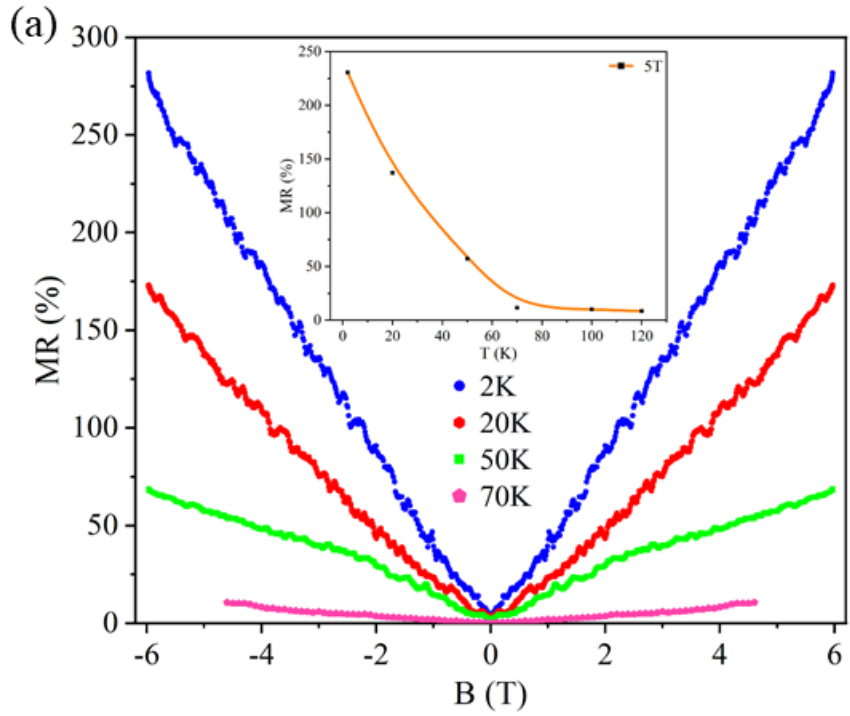

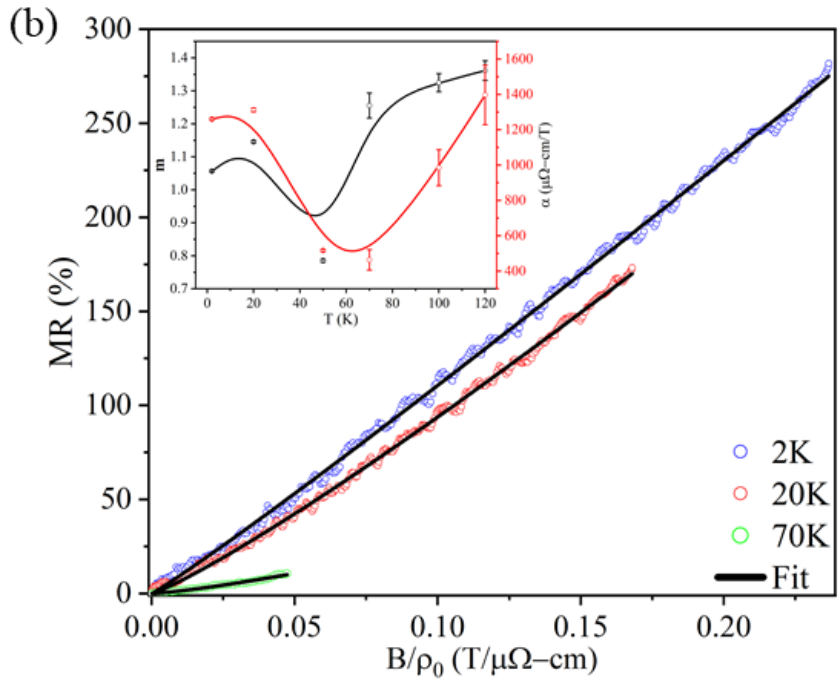

**Figure 4**

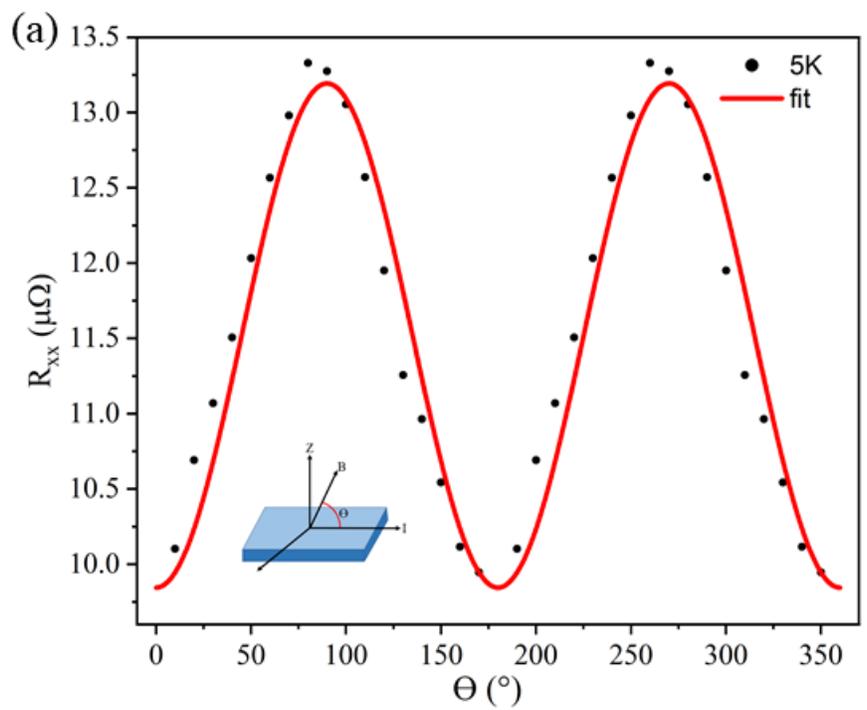

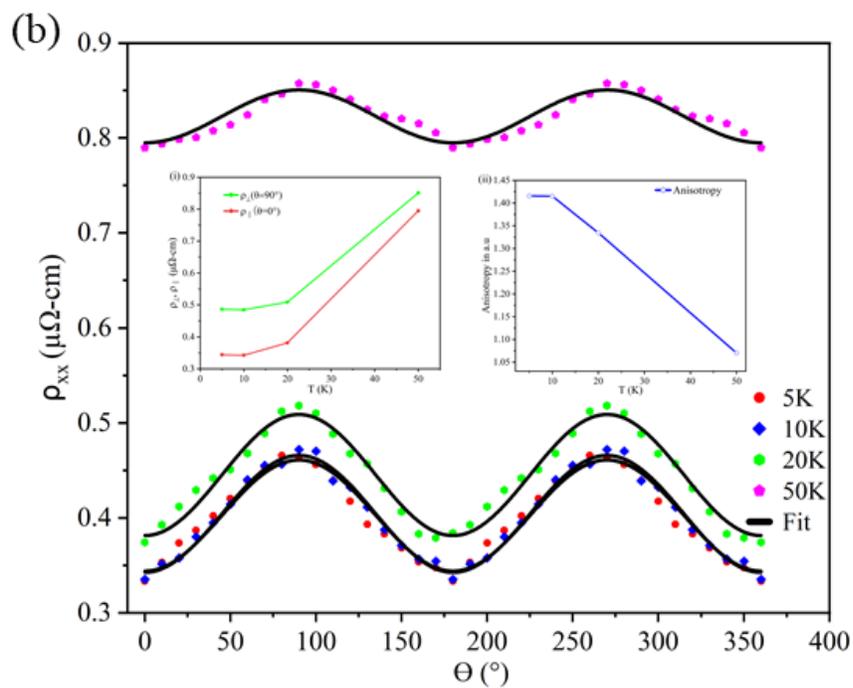

**Figure 5**

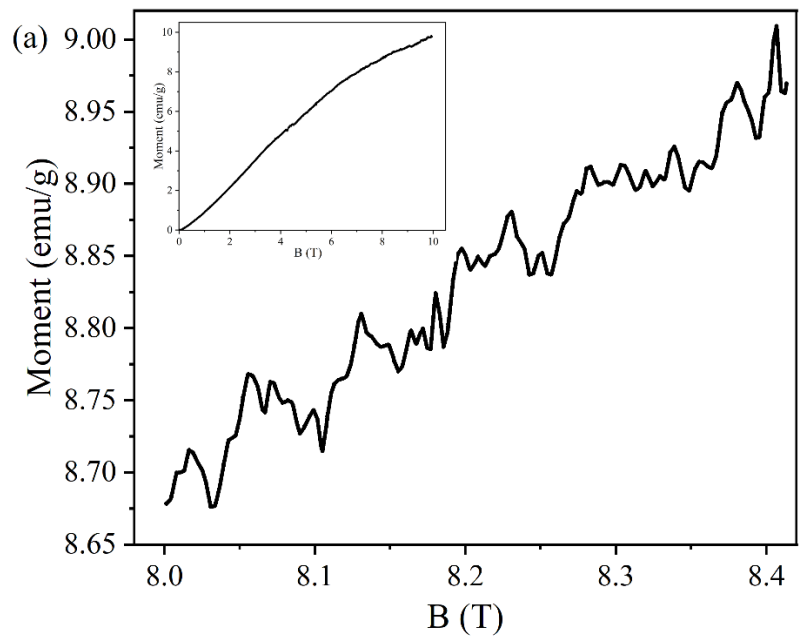

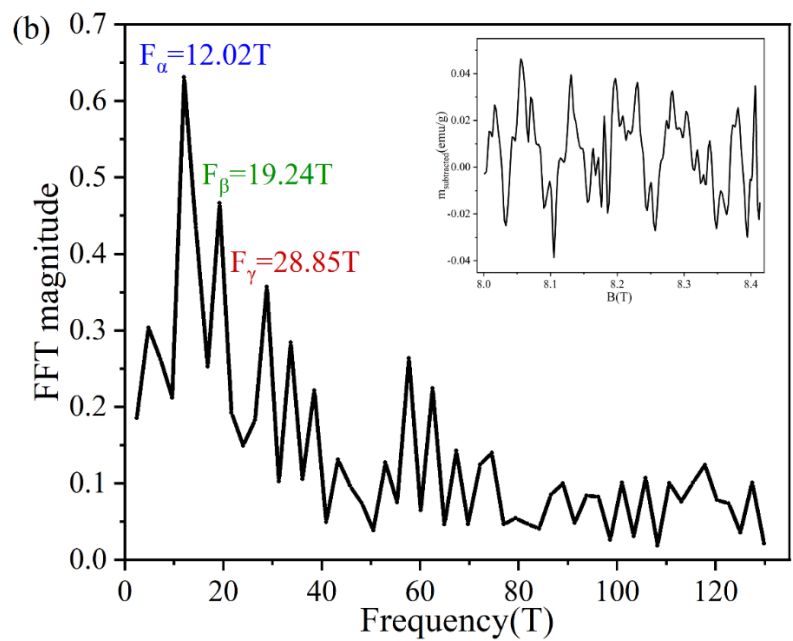

**Figure 6**